\documentstyle[12pt,aaspp4,flushrt]{article}

\newcommand{\baron}{Baron et al. (1995) }
\newcommand{\bartel}{Bartel et al. (1994) }

\newcommand{\fransnusei}{Fransson et al. (1996) }  
  
\newcommand{\houfran}{Houck and Fransson (1996) }  
\newcommand{\mara}{Marcaide et al. (1995a) }
\newcommand{\marb}{Marcaide et al. (1995b) }

\newcommand{\standke}{Standke et al. (1996) }  
\newcommand{\shige}{Shigeyama et al. (1994) }  
\newcommand{\suznom}{Suzuki and Nomoto (1995) }  
\newcommand{\vandyk}{Van Dyk et al. (1994) }

\newcommand{\woos}{Woosley et al. (1994) }

\received{}
\accepted{}

\begin{document}
%
%
\def\sun{M$_{\odot}$ }

\title{Deceleration in the Expansion of SN1993J}

\author{J. M. Marcaide \altaffilmark{1},
A. Alberdi \altaffilmark{2,}\altaffilmark{3},
E. Ros \altaffilmark{1}, P. Diamond \altaffilmark{4},
I. I. Shapiro \altaffilmark{5}, J. C. Guirado \altaffilmark{1},
D. L. Jones \altaffilmark{6}, F. Mantovani \altaffilmark{7},
M. A. P\'erez-Torres \altaffilmark{1},
R. A. Preston \altaffilmark{6}, 
R. T. Schilizzi \altaffilmark{8,}\altaffilmark{9},
R. A. Sramek \altaffilmark{4},
C. Trigilio \altaffilmark{10}, S.D. Van Dyk \altaffilmark{11},
K. W. Weiler \altaffilmark{12}, A. R. Whitney \altaffilmark{13}  
} 

\altaffiltext{1}{Departamento de Astronom\'{\i}a, 
Universitat de Valencia, 46100 Burjassot, Spain; jmm@vlbi.uv.es }
\altaffiltext{2}{Laboratorio de Astrof\'{\i}sica Espacial 
y F\'{\i}sica Fundamental, 
INTA, 28080 Madrid, Spain}
\altaffiltext{3} {Instituto de Astrof\'{\i}sica de Andaluc\'{\i}a,
CSIC, Apdo. Correos 3004, 18080 Granada, Spain}
\altaffiltext{4}{National Radio Astronomy Observatory, Socorro,
NM 87801, USA}
\altaffiltext{5}{Harvard-Smithsonian Center for Astrophysics,
Cambridge, MA 02138, USA}
\altaffiltext{6}{Jet Propulsion Laboratory, California 
Institute of Technology, Pasadena, CA 91109, USA}
\altaffiltext{7}{Istituto di Radioastronomia, Consiglio Nazionale
delle Ricerche (CNR), Bologna 40129, Italy}
\altaffiltext{8}{Joint Institute for VLBI in Europe, 
Postbus 2, 7990 AA Dwingeloo, Netherlands}
\altaffiltext{9}{Leiden Observatory, Postbus 9513, 
2300 RA Leiden, Netherlands}
\altaffiltext{10}{Istituto di Radioastronomia, CNR, Noto 96017, Italy}
\altaffiltext{11}{Visiting Scientist, Physics and Astronomy Department, 
UCLA, Los Angeles, CA 90095, USA}
\altaffiltext{12}{Remote Sensing Division, NRL, Code 7214,
Washington, DC 20375-5320, USA} 
\altaffiltext{13}{Massachusetts Institute of Technology-Haystack
Observatory, Westford, MA 01886, USA}

\begin{abstract}
  
A rarity among supernova, SN~1993J in M81 
can be studied with high spatial
resolution. Its radio power and distance
permit VLBI observations to 
monitor the expansion of its angular 
structure. This radio structure was previously
revealed to be shell-like and to be undergoing
a self-similar expansion at a constant rate. 
From VLBI observations at the wavelengths of 3.6 and 6 cm 
in the period  6 to 42 months after
explosion, we have discovered that
the expansion is decelerating. 
Our measurement of this  
deceleration yields estimates of    
the density profiles of the  
supernova ejecta and circumstellar material
in standard supernova explosion models.

\end{abstract}

\keywords{
supernovae:individual (SN~1993J);
circumstellar matter
galaxies:individual (NGC3031, M81);
techniques: interferometric;
radio continuum: stars;
}
\section{INTRODUCTION}

Supernova SN~1993J in M81 discovered by 
Francisco Garc\'{\i}a of Lugo , Spain  (\cite{RG93}) 
is a type IIb supernova (SN) whose red giant progenitor
probably had a mass of 12-16 
\sun while on the main sequence; at the time of the explosion,
3-5 \sun likely remained in the He core 
and $\stackrel{<}{\sim}$1 \sun
in the He/H envelope (\cite{F93},
 \cite{S94}, \cite{W94}, \cite{I97}).
The first maximum in the 
supernova
optical light curve has been attributed to shock heating
of the thin envelope and the second 
to radioactive decay of $^{56}$Co
(\cite{N93}, \cite{S94}, \cite{W94}). 
Modelling of 
the X-ray emission (\cite{SN95}) also implies
a relatively low mass envelope  due to 
interaction with a binary companion
(\cite{N93}, \cite{W94}). 
\\

The standard circumstellar
interaction model --hereafter standard model or SM--
for radio supernovae (\cite{C96} and 
references therein) suggests that 
the radio emission arises from a shocked region 
between the supernova ejecta and the 
circumstellar material (CSM) that 
results from the wind of the SN's progenitor star. 
More specifically, the SM considers
SN ejecta with steep density profiles 
($\rho_{ej}\propto r^{-n}$) 
shocked by a reverse shock that moves  
inwards from the contact surface  
and a CSM with density profile
$\rho_{csm}\propto r^{-s}$ shocked by a forward 
shock that moves outwards from the 
contact surface ($s$=2 corresponds to 
a steady wind). For $n$$>$5, self-similar 
solutions are possible (\cite{C82a}); the 
radii of the discontinuity surface,  forward 
shock and reverse shock are then related and 
all evolve in time with a power law R $\propto t^{m}$ 
($t$, time after explosion), where $m$=$(n-3)/(n-s)$. 
\\

SN~1993J is the closest SN that is both 
young and radio bright 
(\cite{W96}) and hence  
offers a unique opportunity for the study of its 
radio structure and the test  
of radio supernova models 
(\cite{C82a}, \cite{SN95}). 
\mara found the radio structure 
to be shell-like. Multiwavelength 
radio light curves and high resolution 
radio images of SN~1993J (\cite{V94}, \cite{M95b}, respectively)
established the self-similar nature of the expansion. 
\\

The technique of VLBI can, in principle, determine $m$ 
directly by simply observing the 
angular growth rate of the supernova. 
\bartel and \marb found that $m$=1 was 
compatible with their results to within their respective 
uncertainties. 
In this paper, we present VLBI results for $\lambda$6 cm 
through October 1996 (42 months after explosion), 
combined with those already published 
for $\lambda$3.6 cm (\cite{M95b}), to estimate the 
deceleration in the 
supernova expansion and to infer the density profiles of the 
supernova ejecta and CSM. 
\\

\section{OBSERVATIONS AND RESULTS}

In our $\lambda$6 cm VLBI observations of SN~1993J, 
global arrays formed by the phased-VLA, antennas
in Effelsberg (Germany) and Medicina and Noto (Italy),
and various subsets of the 10-antenna VLBA were used.
For the first 3 epochs (see Table 1)
MkIIIA instrumentation and a recording 
bandwidth of 56 MHz were used
and the data were correlated at the Max Planck Institut fuer
Radioastronomie in Bonn, Germany. For the last 4 epochs, 
VLBA instrumentation and a recording 
bandwidth of 64 MHz were 
used and the data were correlated at the National Radio 
Astronomy Observatory in Socorro, NM.   
The sources 0917+624, 0954+658, and the 
nucleus of M81 were observed as 
calibrators, the first two 
as amplitude calibrators and the nucleus of M81 
both as an amplitude calibrator and, for 
epochs later than June 1996, as a phase calibrator. 
In all cases we analyzed the data 
using DIFMAP (\cite{Shep94}) in a standard way using 
measured system temperatures 
and either antenna-temperature measurements or 
gain-curve information from each antenna as 
initial calibration. For 
0917+624, we obtained brightness maps using 
self-calibration and 
the source structure determined by \standke
as an initial model. 
The calibration correction factors obtained
with the self-calibration of 0917+624
were then applied 
to calibrate the data of SN~1993J and the 
nucleus of M81. A similar iteration was carried out 
using the very compact, VLBI nucleus of M81 and 
those new calibration corrections were 
also applied to the calibration of the data of SN~1993J. 
 \\

We constructed a map 
of SN~1993J for each epoch, using  
a standard process. We used each of the following 
initial models: a point source, a scaled 
model from a previous epoch, and a 
super-symmetrized scaled model 
(obtained by rotating the scaled model 
by $ndeg$, such that 360/$ndeg$ is an integer $n$, 
then rotating by 2$ndeg$, etc.,
adding all the rotated models, and rescaling 
the resulting flux density distribution). 
The total flux density in each  map was checked 
against the light curve of Van Dyk et al. (1994)
and recent VLA measurements. Agreement was found to be
better than 5\% except for two epochs where the discrepancy
was as large as 8\%. The resultant maps were 
virtually independent of the
starting model and are shown in Plate 1. For this display
circular convolving beams with sizes 
proportional to the number of days elapsed since the explosion 
were used (see Table 1).
Such beams permit both a better visualization of 
the self-similar expansion 
(the radio structure remains
similar except for a scale factor) 
and a better estimate of the 
deceleration parameter $m$. 
In Figure 1 we show the map
from the latest epoch (22 October 1996) convolved with an 
elliptical gaussian beam whose half-power size
is given by the corresponding size of the main lobe
of the interferometric beam from that epoch, 
so that the details of the source structure are more
visible than in Plate 1.

Each map of SN~1993J shows a shell-like 
radio source. 
The inferred source size depends on how the map
is constructed and how it is measured. 
Because of the  non point-like size of the VLBI beam, 
a positive bias is introduced in 
the size estimate of each map:  
The estimated size is larger than the true size. 
The fractional bias 
will systematically decrease for a source 
increasing in size if the same 
beam applies for all observations.
If uncorrected, this bias introduces a bias in the 
estimate of the growth rate of the source. 
However, for self-similar expansion, as here, 
a method can be found (see below) such that
the bias can be kept approximately constant with  
source growth and hence does not significantly 
affect the estimate of the deceleration parameter. \\

If the shape of the  expanding source does not change and 
the expansion rate is nearly constant we can largely 
avoid introducing 
a spurious deceleration
by using a beam size proportional to the number
of days between the explosion and the epoch of the source map.
Plate 1 (see also \cite{M95b}) shows that 
we are indeed in such a situation.
An alternate mapping procedure based on
using beam sizes proportional
to actual source sizes 
would produce, in principle, 
a (slight) improvement. In practice, other map 
errors would probably prevent any discernible  
improvement.\\

To use convolving beams as similar as feasible 
to the VLBI  
beams and still use the above-described procedure, we
chose a range of convolving beam sizes so that each is 
always within a factor of two of its VLBI beam. 
For the early epochs, we therefore chose small 
convolving beams and overresolved our images 
by almost a factor of two; for the late epochs, we chose
large convolving beams and 
degraded the map resolution by almost a factor of two. 
We also applied the same criteria to the 
$\lambda$3.6 cm maps (\cite{M95b}). 
\\

In Table 1, we list the measured outer radius of 
the supernova shell for each epoch of observation
and plot the results in Figure 2. 
Each such size was estimated by 
the average of the source diameter
at the 50\% contour level of the map from eight
uniformly-spread azimuthal cuts through
the map. The standard errors quoted in Table 1
result from adding in quadrature an error 
twice as large as the measurement error in the diameters 
and an error in determining the 50\% contour level
location due to map noise and beam size limitations. 
Using R $\propto$ $t^{m}$ for 
the $\lambda$6 cm data yields $m$=0.89$\pm$0.03; 
combining the $\lambda$3.6 and $\lambda$6 cm data 
gives $m$=0.86$\pm$0.02. Figure 2 shows only 
the latter result. The reduced chi-square 
of the model fit is 0.9; the quoted error  
has been scaled to correspond to a reduced chi-square
of unity. In contrast, a fit of a straight line ($m$=1) 
to the data gives a reduced chi-square of 6.0.\\

We find, too, that the better the
calibration and the more VLBI observations at a given epoch, 
the more spherical and the smoother
the resultant image. Thus, some of the 
small emission asymmetries 
in the images may be artifacts.
\\

\section{DISCUSSION}

Our maps give no indication of any structures
developing in the shell by the 
action of either Raleigh-Taylor 
instabilities (\cite{CB95})
or interaction with the CSM. There is also 
no evidence of any departure from circularity as suggested
by some authors to explain the action of a putative binary
companion. We also do not see any emission 
above $\sim$0.5 mJy
from any compact
source at the center of the structure (i.e., 
a pulsar as suggested by \woos and \shige). 
\\

Within the framework of self-similar models, 
measurement of the time dependence of the attenuation 
of the supernova radio emission due to the 
circumstellar plasma allows us to estimate the exponent 
of a power law representation of the 
density profile of the CSM: for free-free 
absorption as commonly invoked in radio supernova models
(Weiler et al. 1996 and references therein), the
opacity, $\tau$, is proportional to the density squared 
integrated along the line of sight. 
Given a supernova radius
R $\propto$ $t^{m}$ and $\rho_{csm} \propto r^{-s}$, 
then $\tau$ $\propto$ $t^{2m(-s+0.5)}$. \vandyk found 
$\tau$ $\propto$ $t^{\delta}$ with 
$\delta$=$-1.99^{+0.38}_{-0.16}$ 
for the
homogeneous component of the CSM. Combining this result
with $m$=0.86$\pm$0.02, 
we obtain $s$=$1.66^{+0.12}_{-0.25}$\,. 
This value
is lower than the $s$=2 in the SM  
for a constant stellar wind, but very close 
to the value $s$=1.7 given by \fransnusei to explain the 
X-ray emission (\cite{Z94}). \vandyk also obtain 
a similar time dependence for the attenuation of a clumpy
medium and hence argue that the clumpy component is
spatially distributed in the same way as the homogeneous
component. \houfran also argue in favor of a clumpy 
medium based on optical line profiles. \suznom  
postulate CSM with homogeneous and clumpy components
to explain X-ray data, but they consider two regions:    
(1) An inner homogeneous region with a density profile
described by $s$=1.7 out to radii smaller 
than $\sim$ 5$\times$$10^{15}$cm and 
(2) an outer clumpy region with 
density profile described by $s$=3 for
the interclump medium at larger radii. Such a model 
of the CSM allows \suznom to fit their model to 
all of the available X-ray data. Specifically, the $s$=3 
clumpy medium is needed to account for the 
hard X-rays and for part of the H$\alpha$ emission. 
The supernova explosion model of \suznom, consisting of
ejecta and a clumpy CSM as described
above, is very different from self-similar models (\cite{C82b}).
\\

The self-similar case with $m$=0.86 and $s$=1.66 gives 
an ejecta density profile of 
$n$=$11.2^{+3.5}_{-1.8}$\,. These values
correspond to steep profiles, indeed much steeper 
than the profiles of white dwarfs ($n$=7), but less steep
than those suggested by \baron from spectral 
analyses or those used by \suznom. 
In the SM for values $n$=11.2 and $s$=1.66, 
the reverse shock radius is $\sim$2\% smaller, 
and the forward shock radius   
$\sim$20\% larger, than the radius of 
the contact surface between shocked supernova ejecta 
and shocked CSM (\cite{C82b}).
\\

\marb estimate that the width of the 
radio shell is about 0.3 times the size of the outer 
radius (or, equivalently, about 40\% of the inner radius).
These authors also estimate expansion speeds 
$\sim$ 15,000 km $s^{-1}$
which are compatible with 
the largest velocities ($\sim$ 11,000 km $s^{-1}$)
measured in H$\alpha$ 
(\cite{F94}, \cite{P95}) 
if (i) the H$\alpha$ emission originates in the 
vicinity of the reverse shock, (ii) a homologous expansion 
is assumed in the ejecta and shocked regions, 
and (iii) the shock
shell is about twice as large as predicted in the SM 
. In an attempt to 
reconcile the SM and the observational
results, \houfran suggest that clumpy ejecta and/or
CSM can broaden the shell. 
On the other hand, the region of the ejecta
shocked by the reverse shock in the model of \suznom
is even larger than that of the CSM
shocked by the forward shock. However, the maximum speeds
of the radio outer shell and of the region of H$\alpha$
emission in the model of \suznom match 
those observed very well, although the density and velocity profiles
in the shell are very different from those of the
standard model.
\\

If we consider only  VLBI results 
from epochs more than 500 days after the explosion, 
we obtain $m$=0.89$\pm$0.03 
and $s$=$1.68^{+0.10}_{-0.27}$\,. 
However, such an age range is in the region in which
the \suznom model suggests $s$=3.0. A contradiction
is apparent and our results therefore 
argue against their model. 
Our estimate of $s$ based on that of $m$ is not
dependent on a given explosion model 
but is a determination from the 
time dependence of the opacity due to an external medium
(\cite{W86},\cite{V94}). Furthermore, such time dependence 
of the opacity has not changed between days 200 and 1000
(Van Dyk, priv. comm.). \\

If the physical picture of the radio and H$\alpha$ emission
in the SM were correct,  
the $\sim$15\% decrease in expansion
speed measured by VLBI between months 12 and 42 after explosion
should be observable in the H$\alpha$ emission.  
On the other hand, if the model of \suznom were correct,
a decrease in the maximum speed of H$\alpha$ 
would not be expected. 
\\

\acknowledgments This research is supported in part 
by the Spanish DGICYT grants PB93-0030 and PB94-1275
and by EU contracts CHGECT920011 and FMGECT950012.
Part of this research was carried out at the Harvard-Smithsonian
Center for Astrophysics under grant AST-9303527
from the National Science 
Foundation (NSF) and at the Jet Propulsion Laboratory, California
Institute of Technology, under contract with the National Aeronautics
and Space Administration. KWW wishes to thank the Office
of Naval Research for the 6.1 funding supporting this 
research. The National Radio Astronomy Observatory is a facility
of the NSF operated under
cooperative agreement by Associated Universities, Inc.

\newpage

\begin{plate}
\caption{ Sequences of radio images of supernova SN~1993J 
at $\lambda$6 cm placed at vertical positions 
proportional to number of days 
elapsed since explosion. The images clearly show
a self-similar expansion. The CLEAN components have been 
convolved with circular beams whose radii are 
proportional to the number of days elapsed since explosion (see text).
The beam sizes have been chosen so as to be within 
a factor of 2 of that of the VLBI  
beam for each observation. Each image in the sequence at left has 
an independent color-coded brightness scale.
In the sequence at right the temperature scale 
is the same for all the images 
(maximum brightness temperature 8.88 mJy/beam).}
\label{plate1}
\end{plate}

\begin{figure}
\caption{ Map at $\lambda$6 cm of SN~1993J from 22 October 1996, 
 1304 days after
explosion. The maximum brightness is 2.48 mJy/beam. 
The elliptical gaussian beam
used in the convolution to obtain this CLEAN map is shown 
in the lower left and has
FWHM of 1.18 $\times$ 0.96 $mas$ with the major axis 
along position angle $5^{\circ}$.     
}
\label{fig1}
\end{figure}


\begin{figure}
\caption{Angular radius of SN~1993J vs. days after explosion. 
The continuous line results from a fit of 
R $\propto t^{m}$ 
(t, time after explosion) to all the data. The value of $m$ 
obtained is 0.86$\pm$0.02. 
For comparison, we show a straight line fit ($m$=1) to all the 
data as a dashed line (see text).
The $\lambda$3.6 cm data come from
Marcaide et al. 1995b.}
\label {fig2}
\end{figure}

\newpage

\def\por{$\times$}
\def\deg{$^{\circ}$}

\begin{small}
\begin{deluxetable}{lccclcccc}
\tablewidth{0pc}
\tablecaption{SN 1993J Sizes\tablenotemark{1}}
\tablehead{
\colhead{Date of}    & 
\colhead{Age} &
\colhead{$\lambda$} & 
\colhead{Flux density\tablenotemark{2}} &
\colhead{VLBI} &  
\colhead{Convolution} &  
\colhead{Shell outer radius\tablenotemark{4}}  \\
\colhead{observation}   & 
\colhead{(days)} &
\colhead{(cm)}      & 
\colhead{(mJy) } &
\colhead{Beam\tablenotemark{3} ~ (mas)}      & 
\colhead{Beam\tablenotemark{4} ~ (mas)}      & 
\colhead{($\mu$arcsec)}}      

\startdata
26-SEP-93 & 182  & 3.6 & 78.5 & 0.55\por 0.46 (13.6\deg)   & 0.26  & 464$\pm$90 \nl
22-NOV-93 & 239  & 3.6 & 57.3 & 0.52\por 0.45 (0.7\deg) & 0.34  & 612$\pm$22 \nl
20-FEB-94 & 330  & 3.6 & 51.0 & 0.53\por 0.44 (6.5\deg) & 0.46  & 824$\pm$90 \nl
29-MAY-94 & 427  & 3.6 & 41.5 & 0.61\por 0.52 (25.8\deg) & 0.60  & 1071$\pm$28 \nl
20-SEP-94 & 541  & 6   & 53.4 & 0.99\por 0.82 (-47.2\deg) & 0.76  & 1202$\pm$30 \nl
23-FEB-95 & 697  & 6   & 44.3 & 1.11\por 0.77 (17.2\deg) & 0.98  & 1567$\pm$33 \nl
11-MAY-95 & 774  & 6   & 41.8 & 1.22\por 0.88 (18.0\deg) & 1.09  & 1701$\pm$39 \nl
01-OCT-95 & 917  & 6   & 32.2 & 1.57\por 1.24 (-27.0\deg) & 1.29  & 2026$\pm$55 \nl
28-MAR-96 & 1096 & 6   & 31.3 & 1.58\por 1.23 (-58.0\deg) & 1.54  & 2301$\pm$72 \nl
17-JUN-96 & 1177 & 6   & 26.5 & 1.37\por 1.11 (30.2\deg) & 1.65  & 2414$\pm$102 \nl
22-OCT-96 & 1304 & 6   & 26.1 & 1.20\por 0.97 (-6.1\deg) & 1.83  & 2639$\pm$100 \nl
\enddata
\tablenotetext{1}{The $\lambda$3.6 cm observations were reported 
by Marcaide et al. (1995b)
but the data have been reanalyzed and the shell outer radius reestimated in a 
way similar to that used for the $\lambda$6 cm data, 
as described in the text.}
\tablenotetext{2}{Total flux density used for mapping.}
\tablenotetext{3}{Major axis $\times$ minor axis (position angle) of elliptical gaussian beam.}
\tablenotetext{4}{As explained in text. Quoted errors are 
estimated one standard deviations.}
\end{deluxetable}
\end{small}

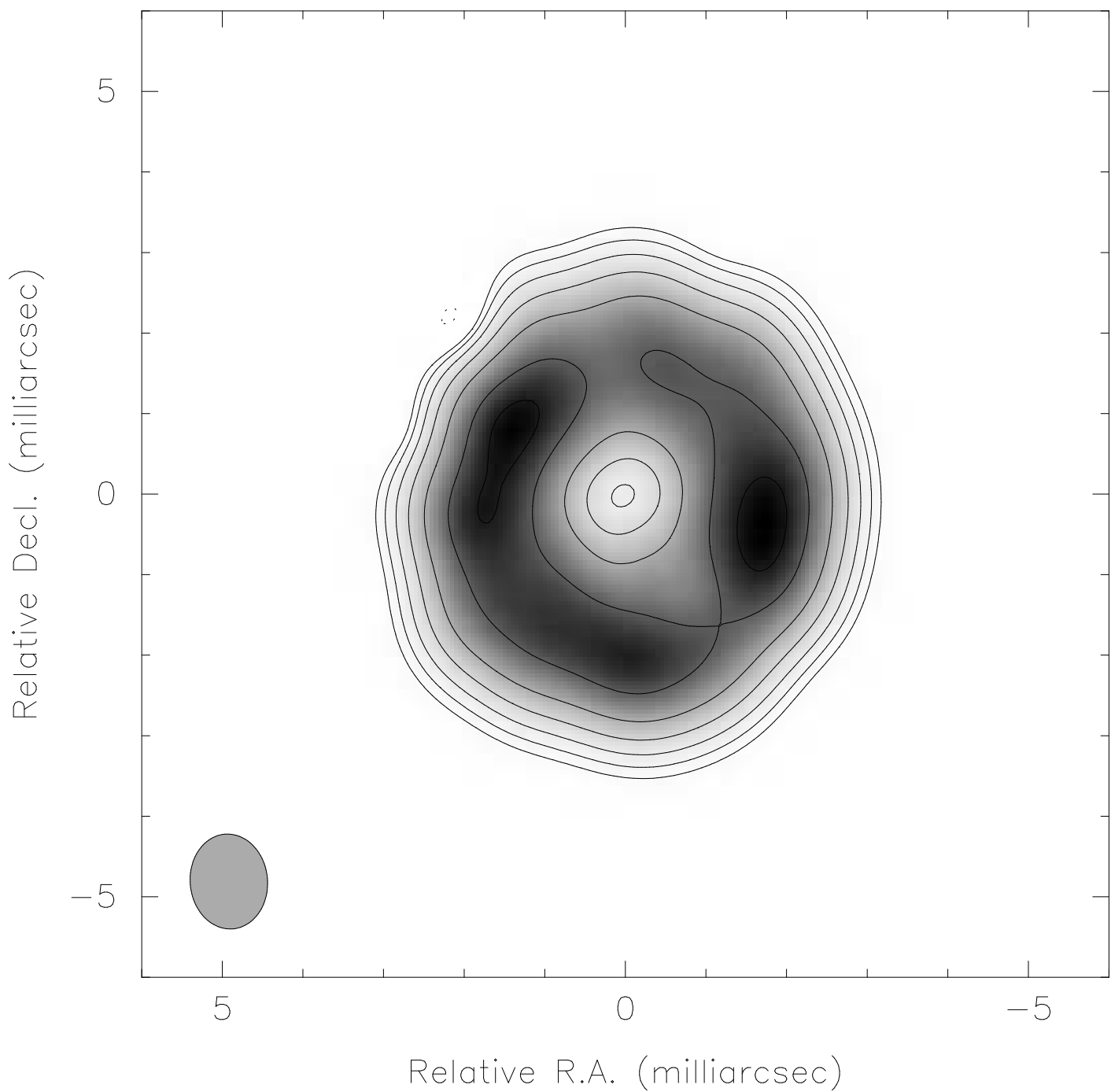
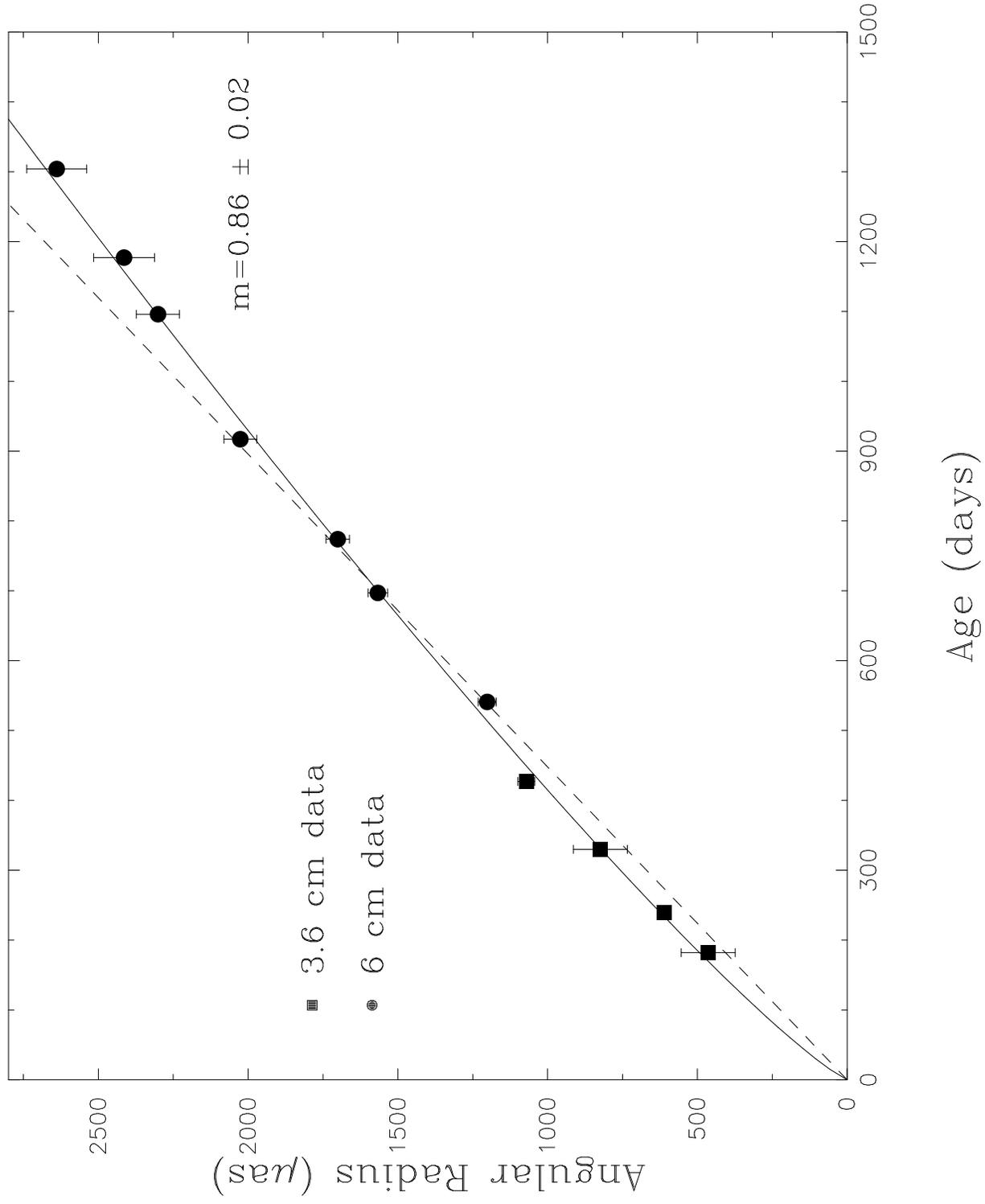
\end{document}